\documentclass{sig-alternate}
\newfont{\mycrnotice}{ptmr8t at 7pt}
\newfont{\myconfname}{ptmri8t at 7pt}
\let\crnotice\mycrnotice%
\let\confname\myconfname%

\usepackage[german,american]{babel}
\usepackage[T1]{fontenc}
\usepackage[utf8]{inputenc}
\usepackage{times}
\usepackage{textcomp}
\usepackage{graphicx}
\usepackage{amssymb}
\usepackage{microtype} 
\usepackage{balance}

\usepackage{url}
\renewcommand{\tt}{\fontencoding{OT1}\fontfamily{cmtt}\selectfont}

\usepackage[usenames,dvipsnames]{color}
\usepackage{booktabs}
\usepackage[numbers,sectionbib]{natbib}
\makeatletter
\def\@copyrightspace{\relax}
\makeatother

\begin{document}

\title{A Two-Level Classification Approach for Detecting Clickbait Posts using Text-Based Features}
\subtitle{The snapper Clickbait Detector at the Clickbait Challenge 2017}

\numberofauthors{6}
\author{
Olga Papadopoulou, Markos Zampoglou, Symeon Papadopoulos, Ioannis Kompatsiaris\\
\affaddr{Centre for Research and Technology Hellas - Information Technologies institute, Greece}\\
\affaddr{\{olgapapa, markzampoglou, papadop, ikom\}@iti.gr}
}

\maketitle

\begin{abstract}
The emergence of social media as news sources has led to the rise of clickbait posts attempting to attract users to click on article links without informing them on the actual article content. This paper presents our efforts to create a clickbait detector inspired by fake news detection algorithms, and our submission to the Clickbait Challenge 2017. The detector is based almost exclusively on text-based features taken from previous work on clickbait detection, our own work on fake post detection, and features we designed specifically for the challenge. We use a two-level classification approach, combining the outputs of 65 first-level classifiers in a second-level feature vector. We present our exploratory results with individual features and their combinations, taken from the post text and the target article title, as well as feature selection. While our own blind tests with the dataset led to an F-score of 0.63, our final evaluation in the Challenge only achieved an F-score of 0.43. We explore the possible causes of this, and lay out potential future steps to achieve more successful results.
\end{abstract}

\section{Introduction}

The number of people turning to social media to get information online, as opposed to traditional news sources such as the websites of news agencies, newspapers and TV channels, has significantly risen in recent years. This has caused a major shift in how news is presented, and has allowed the emergence of new actors in the news market, in the form of outlets that disseminate fake information or rely on ``clickbait'' practices to draw readers' attention.

Clickbait can be defined as the phenomenon where a short post in a social network platform introduces an article by promising exciting or surprising information, but without being sufficiently explanatory for the reader to decide if they are going to be interested in the article or not before clicking. Thus, the article itself may fail to deliver, but the phrasing of the post manages to attract traffic for the site nonetheless. Clickbaiting is thus a marketing technique for attracting readers even in the absence of interesting content.

While not strictly illegal or dangerous, clickbaiting is often frowned upon as dishonest. Facebook in particular is taking steps in reducing the presence of clickbait posts from users' timelines\footnote{\url{https://newsroom.fb.com/news/2017/05/news-feed-fyi-new-updates-to-reduce-clickbait-headlines/}}. Such efforts, however, require a reliable way of automatically detecting clickbait in order to filter results. 

This paper presents our submission to the Clickbait Challenge 2017 \cite{potthast:2017a}, aimed at building a system that detects whether a social media post is clickbait or not based on the post, any accompanying media, and the article itself, including its title, description, keywords, and text. 

Table \ref{tab:sample_clickbait} shows the two main fields for six items from the Challenge dataset, plus their annotations. 

In the rest of this paper we present the methodology we decided to follow and the rationale behind it, the results we achieved using the provided validation data, and the results achieved during the final evaluation. Given the significant discrepancy between the validation results and the actual evaluation, we attempt to explain its causes and plan for more successful future attempts.

\begin{table*}[!t]
\renewcommand{\arraystretch}{1.2}
\caption{Clickbait Challenge examples: post text, associated target article title, and assigned clickbait scores by five annotators. High values correspond to clickbait, while low ones to non-clickbait.}
\label{tab:sample_clickbait}
\centering
\resizebox{\linewidth}{!}{
\begin{tabular}{l| l | l}
\textbf{Post text} & \textbf{Target article title}  & \textbf{Clickbait scores}\\
\hline
SAPVoice: One solution that turns shipping delays to your advantage & SAPVoice: One Solution That Turns Shipping Delays To Your Advantage & [0.33, 0.33, 0.67, 0.67, 1.00]\\ \hline
i thought michelle obama already made this happen? & Escaped Chibok Girl on Anniversary of Abduction by Boko Haram: ``Bring  & [0.00, 1.00, 1.00, 1.00, 1.00] \\ 
& Those Girls Back''  \\ \hline
UK's response to modern slavery leaving victims  & ``Inexcusable'' failures in UK's response to modern slavery leaving victims & [0.00, 0.33, 1.00, 1.00, 1.00]  \\ 
& destitute while abusers go free, report warns  \\ \hline
Panama Papers: Europol links 3,500 names to suspected criminals & Panama Papers: Europol links 3,500 names to suspected criminals & [0.00, 0.00, 0.00, 0.00, 0.33]\\ \hline
Two dead after standoff with cops in Philadelphia @nbcphiladelphia & 2 Men Die in Apparent Murder-Suicide During Barricade Situation in Philly & [0.00, 0.00, 0.33, 0.33, 0.67] \\ \hline
Donald Trump's biggest fan is worried he won't follow through on & Ann Coulter warns Donald Trump of voter backlash if he 'betrays' them  & [0.00, 0.00, 0.33, 1.00, 1.00] \\ 
immigration 'promises' &  on immigration \\
\end{tabular}
}
\end{table*}

\section{Related Work}

The problem of clickbait posts is relatively recent, yet active research is already developing around it. One of the first publications on the subject \cite{chen2015misleading} proposed a set of potential features that could be used for the task, without providing a quantitative analysis of their potential. The proposed features included lexical and semantic features in order to distinguish between high- vs low-quality text by analyzing their stylometry, and  syntactic and pragmatic features to measure the emotional impact of headlines. Besides textual features, they also proposed image and user behavior analysis in order to extract information from the context of the post.

Notable attempts to create clickbait detection systems include an approach using a large number of text features over various classifiers \cite{conf/ecir/PotthastKSH16} and a statistical analysis of the value discrepancies between clickbait and non-clickbait posts over a number of features \cite{journals/corr/ChakrabortyPKG16}. The latter approach also attempts to detect clickbait posts using an SVM classifier, and proposes specialized models using only subsets of features to reflect different user definitions of clickbait. Both approaches rely on various text features, ranging from length statistics and bag-of-words features to common bait phrases.

On the other hand, a different approach to clickbait detection would be to train a deep learning classifier. Two recent approaches have been proposed \cite{journals/corr/Anand0P16,DBLP:journals/corr/RonyHY17}. While both begin by an embedding layer -as is common in neural networks for language processing-, the former uses a Recurrent Neural Network (RNN) layers, while the latter is based on Convolutional Layers.

While we had no prior experience with clickbait detection, we noted a striking similarity of the task to that of fake post detection. While the task of misleading (fake) post detection \cite{castillo2011information}, 
i.e. evaluating whether a post contains true information or not, is not the same as clickbait detection, the approach of extracting text-based features from a post and training a classifier on them is very similar, as are the expectations concerning the distinguishing features for clickbait posts and fake posts. For example, low readability or the increased presence of punctuation are related with both fake and clickbait posts. Given our previous experience with fake post detection \cite{Boididou2017}, we decided to follow a similar approach for our submission to the clickbait detection challenge.

\section{Approach}

Our approach was to use a committee of classifiers, each trained on a different class of text features. The entire list of features was inspired by previous works \cite{Boididou2017,journals/corr/ChakrabortyPKG16,conf/ecir/PotthastKSH16} and is presented in Table \ref{clickbaitfeatures}. Since the task did not include contextual features, either in the form of information on the user account that made the post, or in the form of social media activity around the post (e.g., retweets, replies, etc.), we had to focus on text features from the post and the target.

The bulk of the features were taken from our own previous work on fake post detection \cite{Boididou2017} and was enriched with ideas from \cite{conf/ecir/PotthastKSH16} and \cite{journals/corr/ChakrabortyPKG16}. All features were extracted from the post text and the target title. We also considered extracting features from the target paragraphs and description. However, since not all items in the dataset had values in these fields we decided not to implement an approach for missing values but instead to focus on the fields that were present in all items. Furthermore, we tried to leverage the media information provided by using visual features related to them. Three such features were tried: a binary value indicating the presence or not of media in the post, visual features extracted from the AlexNet FC6 or FC7 activations to see if there are any patterns in the visual content of images that may distinguish clickbait posts, and the presence of ImageNet concepts as detected through GoogleNet (1000 values) to see if there is any correlation between the presence of certain semantic concepts and clickbait. 

Features from our recent work on tweet verification \cite{Boididou2017} include the number of characters, words, question marks, exclamation marks, negative and positive sentiment words, presence of happy and sad emoticon, presence of question and exclamation mark, presence of first, second and third order pronoun, number of slang words, presence of hashtag, and readability score. 
Most of these features are straightforward to extract, while for the most complex of those instructions can be found in \cite{Boididou2017}. 
A second set of features were taken from \cite{journals/corr/ChakrabortyPKG16}, using the provided code\footnote{\url{https://github.com/bhargaviparanjape/clickbait}}. These include the average word length, longest dependency, presence of determiners and pronouns, number of common words, presence of punctuations, whether the text begins with a digit, number of hyperbolic terms, subject of title, percentage of stop words, and presence of word contractions. Finally, we wrote our own extraction code for a number of features, including presence of colon, presence of ``please'', presence of @. 

Number of nouns, number of adjectives, number of verbs, number of adverbs, text voice, Part-Of-Speech (POS) histogram, Named Entity histogram, and sentiment were extracted using the Stanford NER library\footnote{\url{https://nlp.stanford.edu/software/CRF-NER.shtml}}.
Bag-of-words and n-grams were extracted using WEKA\footnote{\url{https://www.cs.waikato.ac.nz/ml/weka/}}.
GID categories were extracted using the General Inquirer\footnote{\url{http://www.wjh.harvard.edu/~inquirer/homecat.htm}}. The post text similarity to the target title, and similarly the post text similarity to the target description were calculated by tokenizing all texts and counting the percentage of shared words. Finally, the visual concepts and visual features were extracted using Caffe\footnote{\url{http://caffe.berkeleyvision.org/}}.

\begin{table}[!t]
\renewcommand{\arraystretch}{1.2}
\caption{A categorization of tested features. Each block was treated as an entire descriptor through concatenation. The presence or absence of media was also used as a descriptor, but since it was a single value, it was concatenated to the other descriptors and not tested on its own}
\label{clickbaitfeatures}
\centering
\resizebox{\columnwidth}{!}{
\begin{tabular}{l|l|l}
\textbf{Morphological (MOR)} & \textbf{Stylistic (STY)} & \textbf{Grammatical (GRA)} \\
\hline
\# characters & \# slang words & POS histogram \\
\# words & Readability score & NE histogram\\
\# question marks & has colon (:) & text voice \\
\# exclamation marks & has ``please'' & is passive voice \\
\# uppercase chars & begins with digit & \# adjectives \\
has question mark & has hashtag &  \# verbs\\
has exclamation mark & has @ & subject of title\\
has 1st/2nd/3rd person pronoun & has contractions & \% of stop words \\
average word length & has punctuations & \\

longest dependency & & \\
has determiners \& pronouns & & \\
\# common words & & \\
text simlarity to target title & & \\
text simlarity to target descr. & & \\
\hline
\multicolumn{3}{c}{} \\
\textbf{Bag-of-Words (BoW)} & \textbf{GID} & \textbf{Sentiment (SEN)} \\
\hline
BoW& 182 categories & \# pos/neg words\\
& & has happy/sad emoticon \\
& &  \# hyperbolic terms \\
& & overall sentiment \\
& & \\
\textbf{Total length: $\sim$1000} &\textbf{Total length: 182} &\textbf{ Total length: 6}\\
\textbf{(depends on training dataset)} & &\\
\hline
\multicolumn{3}{c}{} \\
\textbf{n-grams (NGR)} & \textbf{Vis. Concepts (VIS1)} & \textbf{Vis. Content (VIS2)} \\
\hline
1-4 grams & ImageNet classes & AlexNet FC6 activations \\
  &   & or FC7 activations \\
& & \\
\textbf{Total length: $\sim$2000} & \textbf{Total length: 1000} & \textbf{Total length: 4096}\\
\textbf{(depends on training dataset)} & &\\
\end{tabular}
}
\end{table}

\begin{figure*}[t]
\centering
\hskip-0.15cm
\includegraphics[width=\linewidth]{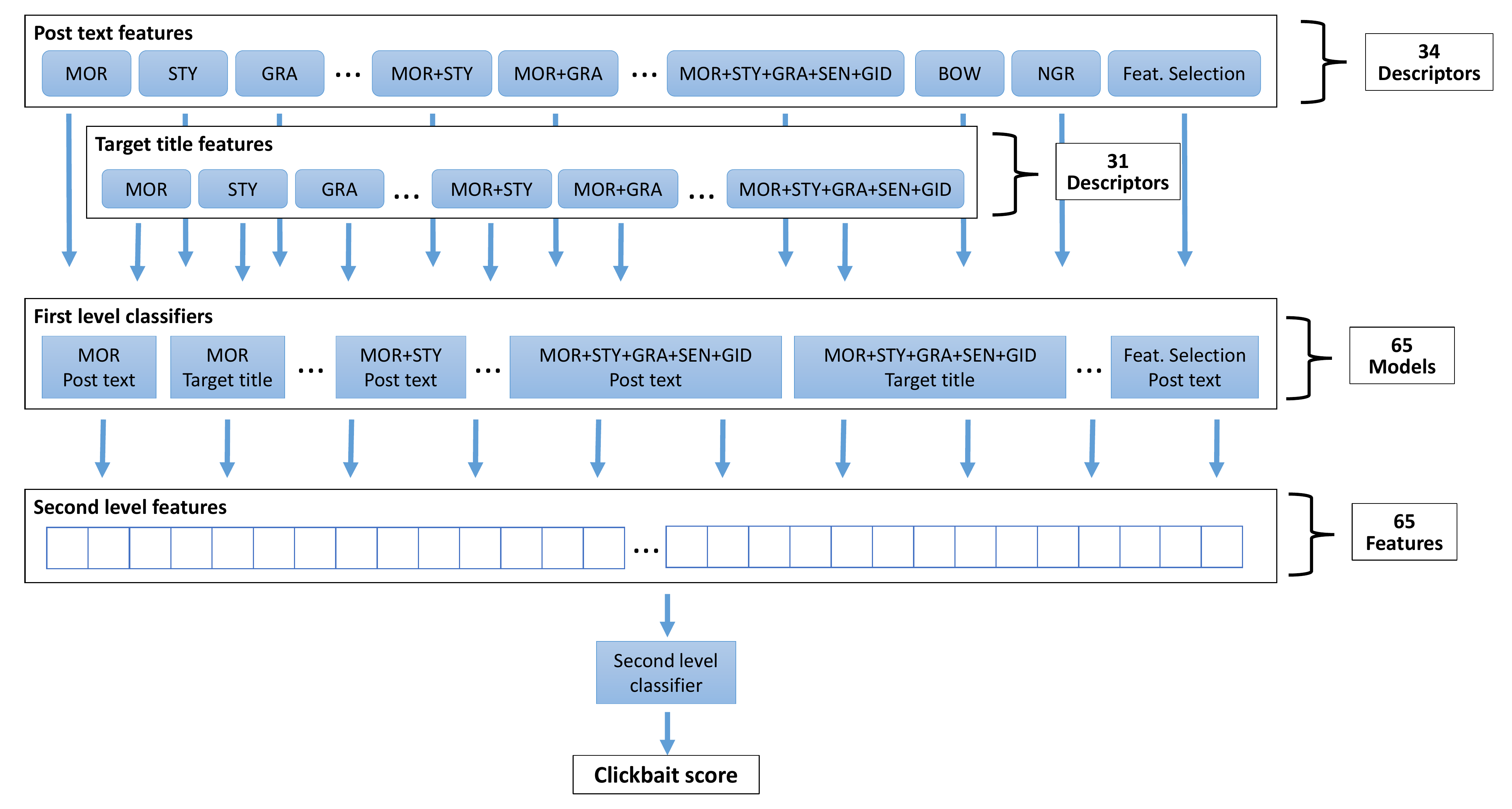}
\caption{The proposed two-level classification architecture.}
\label{fig:architecture}
\end{figure*}

The classification approach was based on a combination of early and late fusion with redundancy, where classifiers were trained on a number of different feature vectors to form a final committee. Text features from the post text and target title were extracted individually. We experimented with both individual features and concatenations of different feature types for the same text source. 
Figure \ref{fig:architecture} presents a visual overview of the proposed classification scheme.
In all experiments, the features were normalized and -following early experiments with a number of different classifiers- a Logistic Regression classifier was used to create a classification model. We further evaluated the impact of feature selection on the concatenated feature vectors. Since these experiments resulted in multiple models with outputs that do not overlap entirely, for second-level fusion we tested both majority voting and building a second-level classifier.

The task datasets were not only annotated with binary tags (1/0: \textit{clickbait}/\textit{non-clickbait}) but also with five quantitative scores from different annotators. 
As can be seen in Table \ref{tab:sample_clickbait}, these scores may differ significantly between annotators, which makes the task much more challenging. We decided to approach the task through the mean of these five values, but it is a consideration that the variance of the values might also have to be taken into consideration.

Motivated by the non-binary provided annotation, we considered two training schemes: one based on binary outputs, resulting from the binarization of the annotation, and another based on a more nuanced class scheme with four classes (corresponding to mean scores in the ranges [0, 0.25), [0.25, 0.5), [0.5, 0.75), [0.75, 1]. However, the latter did not produce promising results, thus we decided to focus on binary classification and indirectly use the output probability of the logistic regression as a metric of how clickbaity an item is.

We used the union of the provided training and validation sets  \cite{potthast:2017b} to randomly create three subsets in order to train and evaluate our classifiers. The entire dataset contained 22,034 items out of which we selected 18,234 at random to use as the main training set (Set A). A second set of 2,000 items was used for testing the first-level classifiers during the preliminary tests, and for training the second-level classifier during the final model design (Set B). Finally, the 1,800 remaining items were used for testing the second-level classifier (Set C). The code of our approach is available on GitHub\footnote{\url{https://github.com/clickbait-challenge/snapper}}.

\section{Evaluation Results}

\begin{table}[!t]
\renewcommand{\arraystretch}{1.2}
\caption{The performance of individual features and feature concatenations extracted from the post text.}
\label{tab:posttext_results}
\centering
\resizebox{\columnwidth}{!}{
\begin{tabular}{l| l l l}
\textbf{Features} & \textbf{F-score}  & \textbf{Precision} & \textbf{Recall} \\
\hline
\textbf{Morph} & 0.47 & 0.71 & 0.35 \\
\textbf{Styl} & 0.22 & 0.67 & 0.13 \\
\textbf{Gram} & \textbf{0.54} & 0.70 & 0.44 \\
\textbf{Sent} & 0.03 & 0.59 & 0.01 \\
\textbf{GID} & 0.36 & 0.66 & 0.25 \\
\hline
\textbf{Morph\_Styl} & 0.51 & 0.73 & 0.40 \\
\textbf{Morph\_Gram} & \textbf{0.58} & 0.72 & 0.48 \\
\textbf{Morph\_Sent} & 0.49 & 0.72 & 0.37 \\
\textbf{Morph\_GID} & 0.53 & 0.70 & 0.42 \\
\textbf{Styl\_Gram} & 0.57 & 0.72 & 0.47 \\
\textbf{Styl\_Sent} & 0.26 & 0.65 & 0.16 \\
\textbf{Styl\_GID} & 0.41 & 0.68 & 0.29 \\
\textbf{Gram\_Sent} & 0.55 & 0.70 & 0.45 \\
\textbf{Gram\_GID} & 0.56 & 0.71 & 0.47 \\
\textbf{Sent\_GID} & 0.39 & 0.65 & 0.28 \\
\hline
\textbf{Morph\_Styl\_Gram} & \textbf{0.59} & 0.73 & 0.49 \\
\textbf{Morph\_Styl\_Sent} & 0.52 & 0.72 & 0.40 \\
\textbf{Morph\_Styl\_GID} & 0.54 & 0.71 & 0.44 \\
\textbf{Morph\_Gram\_Sent} & 0.58 & 0.72 & 0.49 \\
\textbf{Morph\_Gram\_GID} & 0.58 & 0.71 & 0.49 \\
\textbf{Morph\_Sent\_GID} & 0.53 & 0.70 & 0.43 \\
\textbf{Styl\_Gram\_Sent} & 0.57 & 0.72 & 0.47 \\
\textbf{Styl\_Gram\_GID} & 0.58 & 0.72 & 0.49 \\
\textbf{Styl\_Sent\_GID} & 0.44 & 0.67 & 0.33 \\
\textbf{Gram\_Sent\_GID} & 0.57 & 0.71 & 0.48 \\
\hline
\textbf{Morph\_Styl\_Gram\_Sent} & \textbf{0.59} & 0.73 & 0.50 \\
\textbf{Morph\_Styl\_Gram\_GID} & \textbf{0.59} & 0.72 & 0.50 \\
\textbf{Morph\_Styl\_Sent\_GID} & 0.55 & 0.71 & 0.44 \\
\textbf{Morph\_Gram\_Sent\_GID} & 0.58 & 0.71 & 0.49 \\
\textbf{Styl\_Gram\_Sent\_GID} & 0.58 & 0.71 & 0.49 \\
\hline
\textbf{Morph\_Styl\_Gram\_Sent\_GID} & \textbf{0.59} & 0.72 & 0.51 \\
\end{tabular}
}
\end{table}

\begin{table}[!t]
\renewcommand{\arraystretch}{1.2}
\caption{The performance of individual features and feature concatenations extracted from the target title.}
\label{tab:targettitle_results}
\centering
\resizebox{\columnwidth}{!}{
\begin{tabular}{l| l l l}
\textbf{Features} & \textbf{F-score} & \textbf{Precision} & \textbf{Recall} \\
\hline
\textbf{Morph} & 0.31 & 0.58 & 0.21 \\
\textbf{Styl} & 0.25 & 0.55 & 0.16 \\
\textbf{Gram} & \textbf{0.46} & 0.57 & 0.39 \\
\textbf{Sent} & 0.20 & 0.51 & 0.13 \\
\textbf{GID} & 0.31 & 0.54 & 0.22 \\
\hline
\textbf{Morph\_Styl} & 0.35 & 0.59 & 0.24 \\
\textbf{Morph\_Gram} & 0.46 & 0.58 & 0.38 \\
\textbf{Morph\_Sent} & 0.33 & 0.58 & 0.24 \\
\textbf{Morph\_GID} & 0.38 & 0.56 & 0.29 \\
\textbf{Styl\_Gram} & 0.46 & 0.57 & 0.38 \\
\textbf{Styl\_Sent} & 0.31 & 0.54 & 0.22 \\
\textbf{Styl\_GID} & 0.38 & 0.54 & 0.29 \\
\textbf{Gram\_Sent} & 0.46 & 0.57 & 0.39 \\
\textbf{Gram\_GID} & \textbf{0.48} & 0.56 & 0.42 \\
\textbf{Sent\_GID }& 0.35 & 0.53 & 0.27 \\
\hline
\textbf{Morph\_Styl\_Gram} & 0.46 & 0.58 & 0.38 \\
\textbf{Morph\_Styl\_Sent} & 0.36 & 0.58 & 0.26 \\
\textbf{Morph\_Styl\_GID} & 0.40 & 0.56 & 0.31 \\
\textbf{Morph\_Gram\_Sent} & 0.46 & 0.58 & 0.38 \\
\textbf{Morph\_Gram\_GID} & \textbf{0.48} & 0.57 & 0.41 \\
\textbf{Morph\_Sent\_GID} & 0.40 & 0.56 & 0.31 \\
\textbf{Styl\_Gram\_Sent} & 0.46 & 0.57 & 0.38 \\
\textbf{Styl\_Gram\_GID} & \textbf{0.48} & 0.56 & 0.42 \\
\textbf{Styl\_Sent\_GID} & 0.39 & 0.54 & 0.30 \\
\textbf{Gram\_Sent\_GID} & 0.48 & 0.56 & 0.42 \\
\hline
\textbf{Morph\_Styl\_Gram\_Sent} & 0.46 & 0.58 & 0.38 \\
\textbf{Morph\_Styl\_Gram\_GID} & \textbf{0.48} & 0.56 & 0.41 \\
\textbf{Morph\_Styl\_Sent\_GID} & 0.41 & 0.56 & 0.32 \\
\textbf{Morph\_Gram\_Sent\_GID} & \textbf{0.48} & 0.57 & 0.42 \\
\textbf{Styl\_Gram\_Sent\_GID} & \textbf{0.48} & 0.56 & 0.42 \\
\hline
\textbf{Morph\_Styl\_Gram\_Sent\_GID} & \textbf{0.48} & 0.56 & 0.42 \\
\end{tabular}
}
\end{table}

An initial set of experiments assessed the potential of the two image-based descriptors (concept- and content-based). However, both approaches led to very poor results. Combined with the fact that many posts did not contain images at all, we decided to only take into account the presence or absence of images as a binary feature, concatenated with the text features.

The first set of text-based evaluations concerned the individual and combined performance of features using a single-level classifier. The features extracted from the post text and the target title were used to train a Logistic Regression classifier on Set A and evaluated on Set B. The results are presented in Tables \ref{tab:posttext_results} and \ref{tab:targettitle_results}. A number of observations can be made from these first results.

First, it is clear that, in most cases individual features on either source of text are not adequate for classification. The highest F-score in this case is achieved with grammatical features, reaching 0.54 for the post text and 0.46 for the target title. All other features perform significantly poorer when taken in isolation. By looking at the results from the feature concatenations, another observation is that a combination of only two features for the post text (morphological and grammatical) reaches an F-score of 0.58, which is very close to the highest F-score achieved in this way through a concatenation of all five feature classes (0.59). Generally, the performance of grammatical features and all concatenations where it participates is higher than for the rest of features, highlighting the importance of grammatical features in detecting clickbait posts. The same applies for both the post text and the target title, although in the latter case the performance of the classifier is significantly lower.

Overall, the best performance achieved in this manner is 0.59 for the concatenation of all five features extracted from the post text. This is admittedly a rather low score for the task, so further experiments were carried out towards improving the results. 

Another run was based on feature selection on the feature set that worked best, i.e. the five feature concatenation from post text. We used correlation-based feature subset selection in the hopes of improving performance, however this led to a decreased F-score of 0.53 (Precision: 0.69, Recall: 0.43). 

Our next set of experiments were run using 10-fold cross-validation on Set A in order to have a more stable estimate of the feature performance. We performed these evaluations for the concatenation of the five features for four variations: extracted from the post text, extracted from the target title, extracted from both and concatenated, and concatenated from both followed by feature selection. The results are shown in Table \ref{tab:crossval_results}. A clear improvement can be seen for the concatenation of the two sources (post text and target title). Again, feature selection led to deterioration of the overall performance.

\begin{table}[!t]
\renewcommand{\arraystretch}{1.2}
\caption{The performance of individual features and feature concatenations extracted from the target title.}
\label{tab:crossval_results}
\centering
\resizebox{\columnwidth}{!}{
\begin{tabular}{l| l l l}
\textbf{Source} & \textbf{F-score} & \textbf{Precision} & \textbf{Recall} \\
\hline
\textbf{Post text} & 0.52 & 0.92 & 0.36 \\ 
\textbf{Target title} & 0.53 & 0.70 & 0.43 \\ 
\textbf{Both} & \textbf{0.62} & 0.86 & 0.49 \\ 
\textbf{Feature selection} & 0.44 & 0.58 & 0.35 \\ 
\end{tabular}
}
\end{table}

Having trained all these different models, we finally designed a classifier to take advantage of them and evaluated whether we could exploit their potential complementarity to increase the overall performance. The final classifier consisted of 65 individual models, fused in a second-level training scheme. We trained 31 classifiers on the 31 feature vectors listed in Tables \ref{tab:posttext_results} and \ref{tab:targettitle_results}. We used these features extracted from the post text, and 31 more formed from the concatenation of those extracted from the post text and the target title. To these 62 features we added the BoW descriptor extracted and the n-gram feature, both extracted from the post text, and finally the feature-selected five-feature vector extracted from the post text. 65 logistic regression classifiers were trained on these features and a 10-fold cross-validation was run on all 65 models on Set A. Thus, we acquired an estimate for each item and each model, without it having participated in the training set. The 65 outputs were concatenated in a single feature vector, and a second-level logistic regression classifier was trained on all items. Finally, the resulting 2nd-level model was evaluated on Set C. The resulting F-score was 0.63 (Precision: 0.91, Recall: 0.49), and this was the approach we decided to use in our final submission.

For our final submission we followed the same architecture, but trained on a broader base: to get the 1st level outputs, we ran a 10-fold cross-validation of all 65 models on Set A, and at each fold we kept the 65 outputs for all items in the validation set. Then, the second level classifier was trained on all these estimates from Set A. The final submission was evaluated through TIRA \cite{potthast:2014}. However, there was a significant discrepancy in the performance achieved during the classification of our test set, to the one achieved using TIRA. Although our achieved F-score was 0.63, the model performed much worse on TIRA, yielding an F-score of 0.43 (Precision: 0.28, Recall: 0.89). There was a very significant bias towards False Positives, which was the exact opposite behaviour than the one observed in our own evaluations. Furthermore, the feature extraction process and the classification using 65+1 classifiers for each item led to a very slow final model compared to all other submissions. 

Admittedly, an F-score of 0.63 is in itself not satisfactory for any detection task. However, given the final evaluations results, it would have been average compared to other submissions. This may be taken to highlight the difficulty of the task itself, especially when approached with a traditional methodology of extracting text features and feeding them to a classifier. However, our final evaluation output was much lower than that. Barring the presence of an error in the submitted code from our side, this may be indicative of a significant difference between the datasets used for training/testing and the one used for the final evaluation. It may also point to a lack of generalization ability of our own classifier. 

\section{Conclusion}

We presented our submission to the 2017 Clickbait Challenge. The submission was based almost exclusively on text features. The approach used established features from similar approaches in the literature, combined with features from our own experience from fake post detection, plus a number of features specifically tailored for the Challenge. The features were fused using both concatenation and a second-level classifier, in a system which combined features from the post text only and features from the text post concatenated with features from the target title. It was shown that features from the text post gave the best performance (F-score 0.59), and the combination of all features gave a small improvement (F-score 0.63). Despite this performance being on par with other submissions in the Challenge, the final evaluation result on TIRA led to an F-score of 0.43, which may have been the result of overfitting on our training data, despite the measures we took (keeping a separate subset for testing, which did not participate in training). 

Overall, our evaluations showed the limitations of our approach, as even the best F-score of 0.63 is unsuitable for real-world settings. Our extensive evaluations showed that text features alone or in complex combinations could not reach satisfactory performance. 

In our future efforts, we would like to focus more on specifically exploiting inconsistencies between the post text and the target text/title, instead of separately describing both. In our current feature set, the only features that did that were the similarity percentage between the post and target, categorized under ``grammatical'', and it may be no coincidence that these features were the ones that performed best.
We should also put more effort in finding ways to exploit the accompanying data provided for each item, such as the available media and keywords. Our attempts to incorporate visual information did not contribute to the results, so we should reconsider the description approach for images or other associated media. 

In conclusion, despite the unsatisfactory performance of the submitted system, there were many lessons learned during this year's Clickbait Challenge, which may lead to more successful attempts in future editions of the challenge.

\balance
\begin{raggedright}
\bibliography{clickbait17-notebook-lit}
\end{raggedright}
\end{document}